# Towards building a monitoring platform for a challenge-oriented smart specialisation with RIS3-MCAT.

Enric Fuster[1], Tatiana Fernández[2], Hermes Carretero[1], Nicolau Duran-Silva[1,3], Roger Guixé[1], Josep Pujol[1], Bernardo Rondelli[1], Guillem Rull[1], Marta Cortijo[2], Montserrat Romagosa[2]

[1]SIRIS Lab, Research Division of SIRIS Academic
[2]Ministry of Economy and Finance, Government of Catalonia
[3]LaSTUS Lab, TALN Group, Universitat Pompeu Fabra, Barcelona, Spain

**Abstract**

In the new research and innovation (R&I) paradigm, aimed at a transformation towards more sustainable, inclusive and fair pathways to address societal and environmental challenges, and at generating new patterns of specialisation and new trajectories for socioeconomic development, it is essential to provide monitoring systems and tools to map and understand the contribution of R&I policies and projects. To address this transformation, we present the RIS3-MCAT platform, the result of a line of work aimed at exploring the potential of open data, semantic analysis, and data visualisation, for monitoring challenge-oriented smart specialisation in Catalonia. RIS3-MCAT is an interactive platform that facilitates access to R&I project data in formats that allow for sophisticated analyses of a large volume of texts, enabling the detailed study of thematic specialisations and challenges beyond classical classification systems. Its conceptualisation, development framework and use are presented in this paper.

## 1. INTRODUCTION

The challenges posed by globalisation, technology, climate change, and the COVID-19 pandemic require significant changes in our way of living. Although large transition costs are associated with a successful attainment of all those challenges, the potential opportunities brought about are enormous (Bigas *et al.*, 2021). The European Commission aims to accelerate the green transition by implementing the Green Deal (European Commission, 2019) and by allocating funds in the cohesion policy framework and the Horizon Europe (European Commission, 2021) program to mobilise European research and innovation (R&I) ecosystems toward tackling outstanding societal challenges, such as the Sustainable Development Goals (SDGs) defined by the UN (United Nations, 2019). However, reaching these goals requires changes in the forms of cooperation between governments, companies, academia, and other societal stakeholders, new ways of combining knowledge from diverse disciplines, and new tools for evaluating the impact of public policy and R&I (Bigas *et al.*, 2021).

In this same direction, strategies for smart specialisation (S3) (European Commission, 2014), which are dynamic agendas for economic and social transformation based on R&I and articulated through entrepreneurial discovery processes (EDP), are becoming extremely important. Through the EDP, governments, companies, research and innovation stakeholders and civil society organisations and associations collaborate to identify challenges and areas of priority for action, and engage in collaborations towards more sustainable development pathways. In this context, it is therefore key to develop new monitoring systems and tools that help better understand how different actors in the R&I ecosystem are contributing to the SDGs in order to accelerate the transitions, and consequently towards a smarter specialisation.

In Catalonia, smart specialisation is conceived as a forward-thinking open process where innovation stakeholders come together and prioritise challenges and opportunities to be addressed through initiatives, collaborations, policies, and investments (Generalitat de Catalunya, 2022). In the current 2021-2027 programming period, the RIS3CAT monitoring system focuses on understanding how the actions framed in this strategy contribute to:

- articulating sustainable value chains
- promoting business models aimed at generating shared value
- transforming goods and services delivery systems (sociotechnical systems)
- fostering the creation of digital- and technology-based industry
- moving towards a greener, more digital, more resilient and fairer socio-economic model

These transformative processes are complex, as they involve interrelated changes in very different areas (such as the production systems, technologies, markets, regulations, user preferences, infrastructure, and cultural expectations). Accordingly, the monitoring system needs to include and combine different sources of information and types of analysis. Interactive visualisation tools integrating data from different sources are key to identify and analyse emerging areas of specialisation and collaboration networks (within the region and at EU level) in the RIS3CAT priority areas.

Today, EDP, policy implementation and monitoring may be greatly helped by taking advantage of the wider transformative trends in the fields of Open Government and Open Science (European Commission, 2016), which are making more data relevant for the public good increasingly available in open and usable formats (Fuster *et al*., 2020). Data on R&I activities is made available by a series of initiatives. The availability of this data is helpful for the identification of R&D niches and key actors within territorial R&I ecosystems that might be embarked in those transformative processes mentioned above. Simultaneously, the exploitability of this data is increasing due to the advancements in data science, artificial intelligence, and, particularly, in natural language processing techniques, which are being applied to scientometrics to characterise and analyse the textual content of R&I documents. (Fuster *et al*., 2020b).

The European Commission led the way by publishing the CORDIS database of European R&I projects. Since then, public administrations have promoted multiple initiatives such as the European Open Science Cloud, OpenAire, and Zenodo already link projects and their funding with the results they generate (reports, publications, patents, software, etc.). However, the provision and maintenance of open data are highly unequal and do not cover the full range of needs of public policymakers. The availability of open data with sufficient granularity and richness remains a challenge, although it is becoming less limiting as science and technology databases grow in number, size, coverage, quality, interconnection, and content richness. Some major challenges faced at a policy level arise because many of those data sources are not openly available which undermines the participatory processes). Additional challenges include their lack of interoperability in terms of data classification schemes, their institutional identification limiting transversal analyses and their lack of accessibility to non-expert users.

In this context, a line of work has been established to explore the potential of integrating open data, semantic analysis and data visualisation with the aim of developing methodological proposals for monitoring smart specialisation. This exploration, which tackled the challenges linked with the definition of indicators for monitoring emerging areas, territorial patterns of specialisation and collaboration dynamics between different stakeholders and areas of knowledge, led to the development of the RIS3-MCAT interactive platform[1], whose conceptualisation, development framework and use are presented in this paper.

This paper is organised as follows. Section 2 introduces policy objectives and main functions of the RIS3-MCAT monitoring platform. Section 3 presents the five-year co-design and development process. Section 4 presents data sources, system architecture and presents main results of components and features. Section 5 illustrates the main use cases. Finally, Section 6 draws conclusions and recommends future work directions.

## 2. POLICY OBJECTIVES AND MAIN FUNCTIONS

The RIS3-MCAT Platform is an interactive tool aimed at visualising, exploring and analysing the specialisation and collaboration patterns of R&I projects financed with European funds in Catalonia. It is an open government, artificial intelligence and data visualisation project that integrates and makes openly accessible and interoperable data from science and innovation projects, with the aim of contributing to the following objectives:

- understanding the impact of European funds on the specialisation of the R&I ecosystem of Catalonia,

- identifying opportunities to maximise the collective impact of R&I in Catalonia, based on synergies and the coordination of efforts

- providing new evidence that facilitates decision-making by stakeholders in the R&I ecosystem of Catalonia, promoting new dynamics of collaboration and inspiring new public policies;

- raising the profile of Catalan public and private actors that participate in R&I European networks;

- understanding the contribution of European funds to innovative responses to regional priorities, emergent thematics and the Sustainable Development Goals (SDGs).

Apart from the interactive visualisation tools, RIS3-MCAT provides all its curated and enriched data as open data, via dump downloads and via a 5* open data SPARQL Console.

This has facilitated the elaboration of several complementary policy monitoring as well as transversal and thematic analytical reports, published under the "Monitoring RIS3CAT" document collection[2].

*Figure 1: Screenshot from RIS3-MCAT monitoring platform - Network view.*

---

[1] Available at: https://ris3mcat.gencat.cat/.
[2] Available at: https://fonseuropeus.gencat.cat/ca/ris3cat/2030/monitoratge/index.html#googtrans(ca|en)

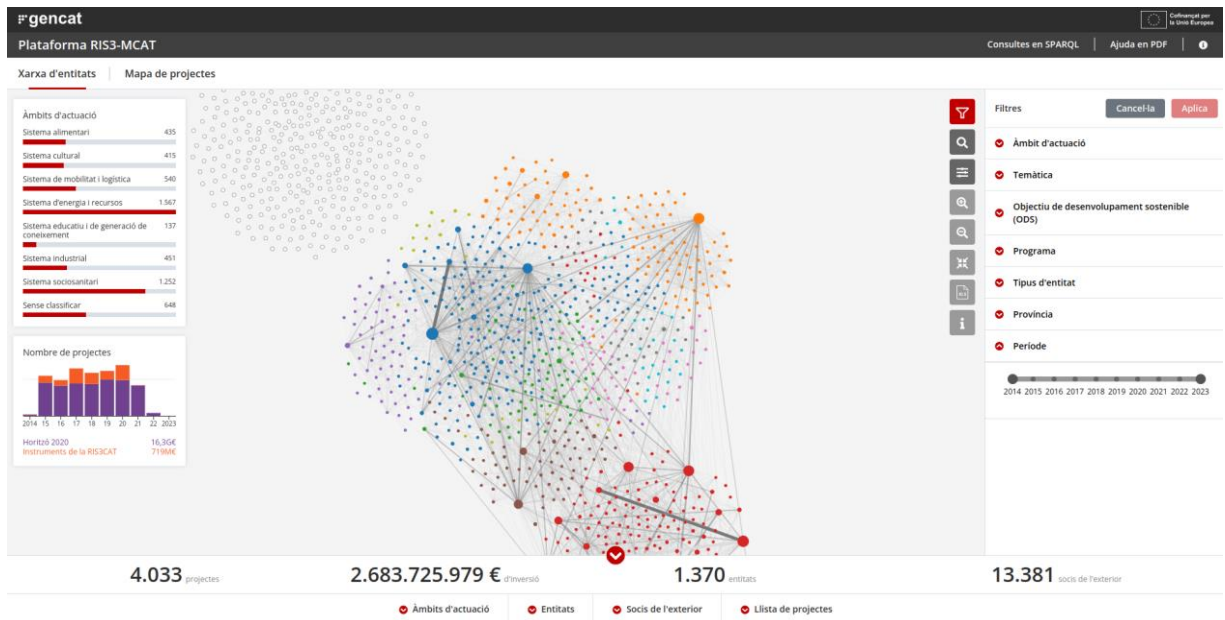

## 3. A FIVE-YEAR CODESIGN AND DEVELOPMENT PROCESS

As of March 31st, 2023, more than 5,000 unique users from 73 countries have accessed the platform, with an average session duration of over 4 minutes. RIS3-MCAT is currently undergoing a major redesign, with new functionalities being introduced to facilitate R&I portfolio analysis, an essential requirement in the new European programming period (2021-27). The new version was published in May 2023. This represents a further advancement in a comprehensive, ongoing process of co-design, review, evaluation, and iterative development that began in 2017, as outlined in the timeline below:

- 2017: Requirement analysis and feasibility study which focused on data availability and quality, as well as prospective front-end and back-end technologies and solutions.

- 2018: Proof of concept development, based on a wide co-design process within the R&I related departments of Generalitat de Catalunya. This led to the first officially published live version, with manually classified Horizon Europe and RIS3CAT R&I projects, focusing on S3 priority analytics and collaboration networks.

- 2019-2020: Consolidation of the proof of concept into a fully finished product. Development and inclusion of the SDG project classification. First development of an analytical report, based on the RIS3CAT taxonomy and an automatic identification of main themes via topic modelling (machine learning), to support the decision-making around the evolution of the RIS3CAT regional priorities for the new programming period. Provision of data for the S3 monitoring official reports. Participatory development of thematic analysis in three domains: Circular Bioeconomy, Artificial Intelligence, and Plastic Waste Reduction.

- 2021: Participatory review and requirement analysis with Catalan R&I stakeholders, leading to the development of new functionalities. More specifically, supporting the identification and analysis of international and inter-regional collaboration and improving the bulk download of the underlying data.

- 2022-May 2023: Adaptation to RIS3CAT 2021-2030, the new policy framework, by automatically reclassifying all projects into the new RIS3CAT regional priorities via deep learning classifiers. Major review and redevelopment of the platform, updating the front-end technology, and streamlining the design patterns. First integration of an emergent classification, the "Topics" based on topic modelling (deep learning). Integration of the first Horizon Europe projects. Addition of the "Thematic project mapping" platform view, which presents all RIS3MCAT R&I projects in a single visualisation, organised by semantic similarity. With this being the first new view introduced in the platform since 2018, it opens the door for further possibilities, such as purpose-built benchmarking tools or geographical representations on a map.

## 4. TOOL OVERVIEW

### 4.1. Data sources

**CORDIS**. Data and metadata related to R&D projects and related organisations which have received funding by the European Commission under the H2020 and FP7 framework programs. The data are accessible through Open Data licence, and updated monthly, provided in the format of CSV, XML and Linked Open data on the CORDIS website. We collect the CORDIS records from UNiCS (Giménez, 2018), an open data platform based on semantic technologies for science and innovation policies which include data cleaning and improved geographical identifications of participants, which are not always correct in the original datasets.

**SIFECAT.** The information system used by Generalitat de Catalunya to manage ERDF operation, and the main data source for regionally-managed R&I projects. There is an initial manual review and improvement by Generalitat officials, notably in terms of institutional naming and geolocalisation, but some transformations have to be derived to fit the data integration ontology and the front-end requirements.

### 4.2. RIS3-MCAT architecture

**Data integration & cleaning**. CORDIS records and projects funded by the Catalan region and ERDF are integrated into different database schemas that are then homogenised into a unified structure with a domain ontology[3]. Access to the data is done through the ontology by means of the Virtual Knowledge Graph system Ontop (Calvanese *et al.*, 2017), which translates input queries formulated over the ontology into executable queries formulated over the underlying database. In order for these different datasets to be integrated properly, the project beneficiaries must be given an identifier that homogenises differences in spelling. The goal is that a same organisation must have the same identifier, even if its name is written differently in each dataset. To this effect, a process of semi-manual disambiguation had been performed by our experts, using the OpenRefine tool[4]. Additionally, for SIFECAT data, each beneficiary had been annotated with the corresponding type of organisation.

---

[3] Ontology schema available here: https://s3-eu-west-1.amazonaws.com/ontology-documentation-ris3cat/index.html
[4] https://github.com/OpenRefine/OpenRefine

**Automatic classification of projects according to the regional priorities.** This approach allows us to capture "top-down" domains when a deeper understanding of the regional dynamics within a specific research area that was previously defined is needed. We took advantage of pre-trained language models for training textual classifiers based on title and project description, one per each of the seven priority "systems" or areas of application by the RIS3CAT 2030 (Generalitat de Catalunya, 2022). A training set was built for each domain and annotated based on weak-supervision and active learning paradigms, which allow weak annotation of projects from a general sample of R&I projects from both FP7 and H2020 frameworks based on some of their metadata[5]. We utilised the tool Argilla (Vila & Aranda, 2023) for annotation and label improving, which allowed keeping interactive feedback from experts to improve label quality. Our best models were based on Specter (Cohan *et al.*, 2020), and it was implemented with the Hugging Face Transformers library (Wolf *et al.*, 2020). From an evaluation on a sample of 500 Catalan projects, predicted labels were compared with human labels from 2 experts, obtaining a macro-averaged f1 of 88.1% of accuracy.

**Topic modelling.** Topic Modelling (TM) is an unsupervised classification problem in machine learning that aims at *discovering* the unknown topics linked with a specific collection of texts, presenting a "bottom-up" picture of the thematics tackled within a specific R&I community. This component takes the project title and abstract, encodes the semantic representation of them based on the Specter sentence-transformer model (Cohan *et al.*, 2020) which was pre-trained on scientific documents, and from clustering vectorial representation of documents based on k-means, we obtained groups of similar projects. Each document was linked to a cluster, and the number of clusters was decided by qualitatively selecting the best trade off between the semantic "richness" of the topics and the overall number of topics (in order not to have neither too large topics nor too little ones) on different runs. Names of clusters were added manually based on keywords frequency and by exploring samples of projects.

**Sustainable Development Goals classification (SDGs).** We identified SDG-related research by using a collection of SDG keywords (a controlled vocabulary) openly available in Zenodo (Duran-Silva *et al.*, 2019), based on a hybrid approach that uses automatic methods for enriching human-crafted keywords. R&I projects are tagged with VocTagger tool[6] on their title and project description.

**Web front-end.** RIS3-MCAT UI and its visualisations were implemented following the W3C standards and using a combination of HTML, CSS and JavaScript, also exploiting third-party libraries (e.g., the D3.js and React.js libraries). The projects and participants information data enrichment was retrieved by querying the SPARQL endpoint.

*4.3. RIS3-MCAT features*

The RIS3-MCAT front-end is composed of four main parts: navigation bar, main visualisation canvas, operations toolbar and statistical modules. Each of the parts is fully reactive to the user interactions. The main visualisation canvas has 2 different representations of the data (that can be explored by the navigation bar), a collaboration network of institutions, and a semantic cartography of projects.

---

[5] Taking advantage of: EC Area, ERC Panel, EC programme, Topic, and Field of Study.
[6] https://github.com/sirisacademic/VocTagger

**Search & Filters.** To facilitate data exploration, the platform offers users various filtering and search options, allowing them to generate customised visualisations. Initially, the platform displays all the integrated data. However, this visualisation can be restricted to subsets of data by applying filters and search parameters or by directly manipulating the network. Filtering features include search by: keyword, participant name, institution type, year province, instrument and programme name, area of action, emerging topic, and SDG. Search features include search by participant and project search based on keyword or text search on title and abstract. All filters are multi selection filters, and all filters work in combination with searches to allow users to define a set of exploration of interest.

**Network analysis.** The network of participants shows the collaboration of Catalan R&I actors. Each node of the network represents a R&I actor with its legal headquarters in Catalonia, and the size of the node is proportional to the volume of the entity's investment in the projects. When two entities collaborate on projects, the nodes that represent them are joined with a line. The size of this line is proportional to the number of projects they share. Force directed graph (network) show the relationships between participants and are defined by the collaborations on projects.

**Semantic map of projects.** The R&I projects are displayed on a 2D canvas, which is organised based on the semantic similarity between them. This setup creates a 'topography' of the R&I activity, making it easier to identify similar projects. It also provides a visual representation of how closely related different themes are, as well as the overlap and connections between different classifications. It is a T-SNE dimensionality reduction of the embeddings obtained in the topic modelling module, and its clusters and names.

**Analytical/Statistical modules.** The information modules, presented at the bottom of the tool, extend the textual and statistical information of projects and participants and their classifications, displaying distributions and relationships. They are reactive to the filtering operations. The present types are: summary indicators, rankings of participants, external partners, and projects table view. Different project information modules show extended information of the projects, and they are available from different parts of the application.

**Data download & SPARQL Endpoint.** The platform offers the possibility of downloading the data filtered interactively as a CSV file, or of making queries about all the data included using SPARQL. Data download is available (XLS format) for the current state of the project's and participants visualisations as well as the regional partners and international partners in the statistical modules.

## 5. ILLUSTRATIVE USE CASES

We have identified interesting scenarios of use by different target users/actors in the territory, with short descriptions for the four illustrative use cases.

- **Use case 1: Search by actors and projects in similar topics (search for expertise and possible collaborations).** Target users for this use case could be R&I stakeholders, such as researchers or private companies.

*Figure 2: Screenshot from RIS3-MCAT monitoring platform - Use case 1*

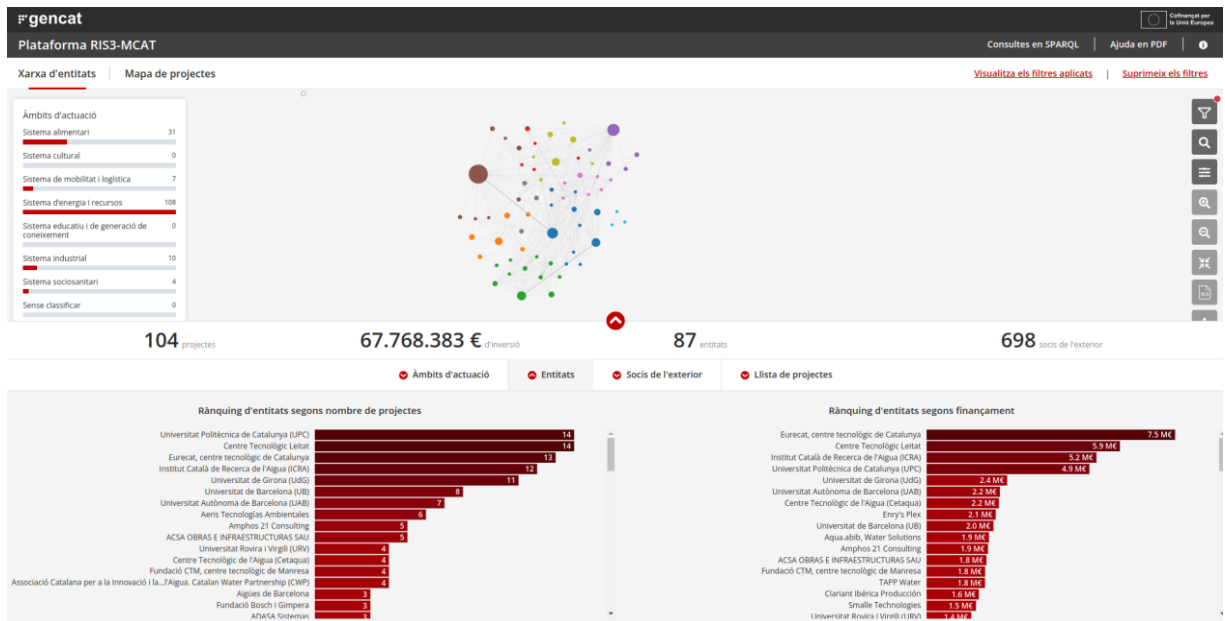

- **Use case 2: Collaboration network for the identification of actors within a thematic and geographic scope.** Target users for this use case could be a R&I policy-makers, with transversal, or thematic / territorial responsibilities.

*Figure 3: Screenshot from RIS3-MCAT monitoring platform - Use case 2*

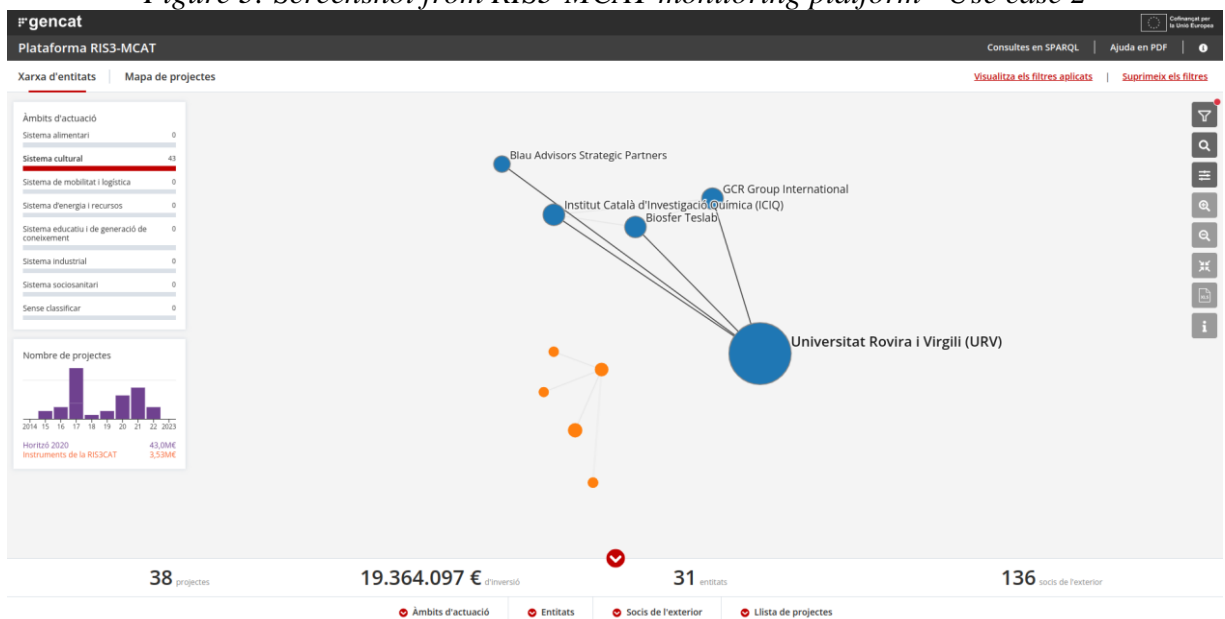

- **Use case 3: State of research for priority area.** This use case allows the study of the intersection between the top-down priority area and emerging topics, in order to find projects in the "core" of the priority, as well as those that are more interdisciplinary/intersectoral. The target user for this could be a leader of a policy or an initiative focused on R&I challenges. This person would use systems thinking methods to thoroughly investigate and address societal priorities or challenges. We can find in the "core" of the priority projects such as "METROFOOD-RI Preparatory Phase Project" or "Connecting the dots to unleash the innovation potential for digital transformation of the European agri-food sector"; and, in the periphery, projects like "Empowering consumers to PREVENT diet-related diseases through OMICS sciences" in health domain, or "Advanced Multi-Constellation EGNSS Augmentation and Monitoring Network and its Application in Precision Agriculture" in satellite navigation. Figure 4 captures an example of this use case.

*Figure 4: Screenshot from RIS3-MCAT monitoring platform - Use case 3*

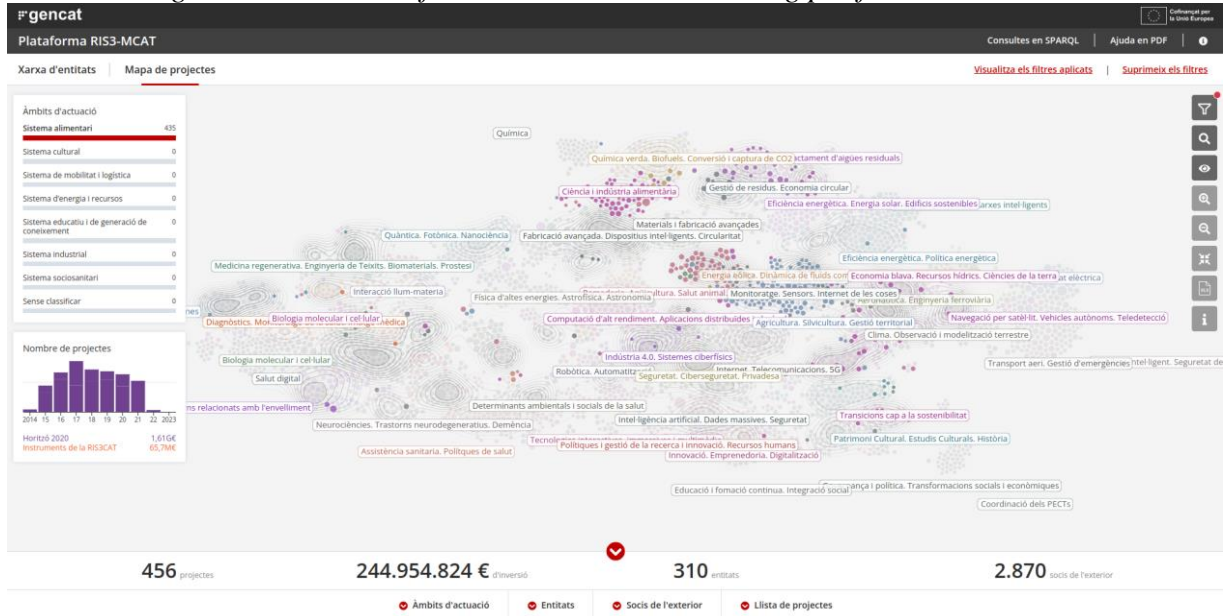

- **Use case 4. Collaboration promotion.** This use case allows the identification of current partners in other countries/regions (and their counterparts in Catalonia) by topic. The target users are mainly internationalisation policy-makers.

*Figure 5: Screenshot from RIS3-MCAT monitoring platform - Use case 4*

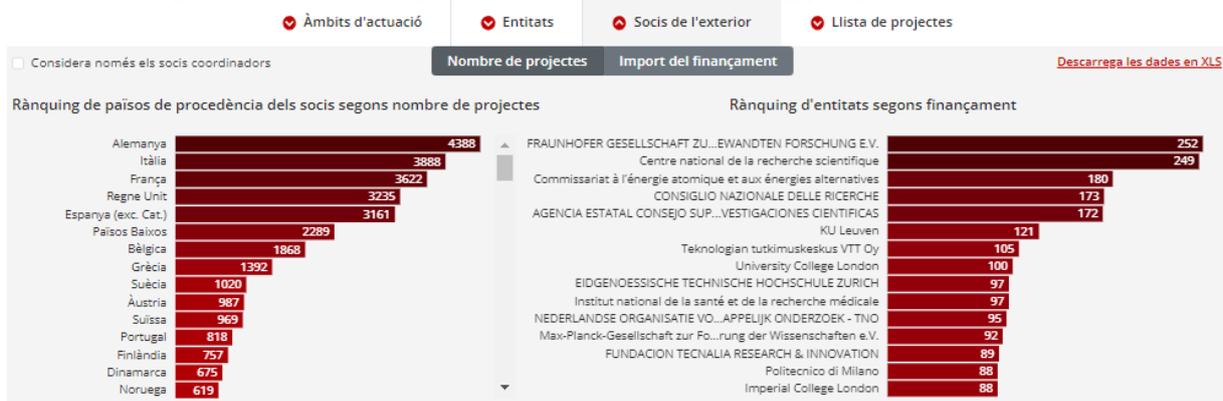

## 6. LESSONS LEARNED, CONCLUSIONS AND FUTURE WORK

This document presents the RIS3-MCAT mapping and monitoring platform, which has tackled the conundrum of monitoring challenge-oriented smart specialisation with available technologies and open data. The development has been shaped through a collaborative design approach, evolving the methods, techniques, and priority areas of focust to meet the needs of policymakers and the strategic and operational demands of quad-helix stakeholders.
The main technical innovations have been done through the use of artificial intelligence and NLP for treating, classifying and semantically enriching large numbers of R&I documents, and the development of interactive exploratory visualisation tools as an entry-point to complex and highly-intertwined data-sources.

Some of the key lessons learned during the process, as well as open challenges and development opportunities for further consideration are as follows:
- A successful and enduring co-design process hinges on strategic adaptability, consistent and high-quality dialogue between stakeholders and providers, and decisive leadership, especially when making bold shifts or redefining project goals. Otherwise, the process may remain too close to the initial idea design, and becomes irrelevant or outdated in the mid and long term.
- Although data gathering, integration and enrichment processes can be automated to a large degree, a certain amount of manual attention and curation will always be necessary. This is due to changes in the policy instruments and data structures, as well as fringe cases and mistakes in the data. Incorporating new projects and data sources is essential to keep the R&I landscape up to date in the monitoring platform.
- There is still relevant S3-related publicly-funded R&I activity not in the platform. The introduction of additional sets of R&I projects, both funded through open calls and through direct public-public agreements, could be an area of focus for future development.
- Apart from the web platform, RIS3-MCAT is an essential source of information for S3 periodic monitoring reports and other ad-hoc analysis. Nevertheless, although publicly accessible the datasets have not been reused beyond the narrow group of RIS3CAT related policy-makers and actors. Therefore, re-publishing the data in portals or repositories with a wider (international) audience could facilitate uptake by new users (notably researchers and policy-officers elsewhere). This would be further enabled by better documentation and sharing of the data enrichment methodologies and pipelines.
- The visualisation and interaction with the Collaboration network and the Semantic map of project features, remains challenging for new users, due to the very large number of entities and their relationships. New facets, zoom-in features, or simplified landing versions could be developed to ease the first impression of complexity.
- During the short timespan of the project, artificial intelligence and natural language processing has advanced very fast, providing new improved tools that have replaced manual thematic classification and offered new, useful, fine-grained classification systems. Also, new user requirements and visualisation ideas have also emerged. To remain relevant in this changing context, monitoring platforms like RIS3-MCAT must keep a flexible part of the budget allocation to allow for exploration, testing and development of new or improved visualisations, functionalities, and analytics.
  - Some of the backlog of ideas regarding data visualisation and front-end functionalities are: incorporating a geographic map, KPI analytical dashboard, European benchmarking tool, and interregional collaboration analysis.
  - Regarding classification and semantic enrichment, it would be interesting to explore new ways of identifying thematic niches and notably, non-thematic functional classifications, such as the five dimensions of socio-technical systems[7] (Geels, 2002; Geels, 2004) which have to be tackled through transformative innovation policy. The addition of these new classification systems would facilitate a wider range of mapping, monitoring and coalition-building efforts.

---

[7] These dimensions are: Science, Technology and Infrastructure, Policy and governance, Investment and finance, Society and culture, Markets.

**Open science practices**
This platform aims at promoting open science policies, allowing the exploration R&I activities in the region and encouraging collaboration. Our project itself is based and generates open data, and we have made intermediate reports about our codesign and lessons learned, publicly available. In our exploration and data integration, we have used open data sources and all data generated is available via SPARQL and CSV formats. Additionally, we have used open source technologies available on GitHub and the data of the platform is under a CC0 licence. Tye SDG vocabularies and VocTagger, partially developed in this context, are available on GitHub and Zenodo. Open science practices are crucial to advance R&I, but also public policies. For this reason, this article is intended to explain, formalise, and communicate the results, decision, and process of this research, so that other actors and regions can benefit in their own initiatives.

**Acknowledgments**
Supported by the Industrial Doctorates Plan of the Department of Research and Universities of the Generalitat de Catalunya. This work was co-funded by the EU HORIZON SciLake (Grant Agreement 101058573).

We acknowledge support of this work by Sergio Martínez (Generalitat de Catalunya), Francesco Massucci (SIRIS Academic), Arnau Quinquillà (SIRIS Academic), Xavi Giménez (SIRIS Academic), and Simon Larmour (SIRIS Academic).

**Competing interests**
This article is authored by the key responsibles of RIS3MCAT at Generalitat de Catalunya alongside its private sector providers (SIRIS Academic). We believe we do not have competing interests.

Transformers: State-of-the-Art Natural Language Processing [Conference paper]. 38–45. https://www.aclweb.org/anthology/2020.emnlp-demos.6